# Towards 2+1 photon tomography: Energy-based selection of two 511 keV photons and a prompt photon with the J-PET scanner.

Masełek R.[1], Krzemień W.[2], Klimaszewski K.[1], Raczyński L.[1], Kowalski P.[1], Shopa R.[1], Wiślicki W.[1], Białas P.[3], Curceanu C.[4], Czerwiński E.[3], Dulski K.[3], Gajos A.[3], Głowacz B.[3], Gorgol M.[5], Hiesmayr B.[6], Jasińska B.[5], Kisielewska D.[3], Korcyl G.[3], Kozik T.[3], Krawczyk N.[3], Kubicz E.[3], Mohammed M.[3, 6], Pawlik-Niedźwiecka M.[3], Niedźwiecki S.[3], Pałka M.[3], Rudy Z.[3], Sharma N.G.[3], Sharma S.[3], Silarski M.[3], Skurzok M.[3], Wieczorek A.[3], Zgardzińska B.[5], Zieliński M.[3], Moskal P.[3]

[1] Department of Complex Systems, National Centre for Nuclear Research, 05-400 Otwock-Świerk, Poland
[2] High Energy Physics Division, National Centre for Nuclear Research, 05-400 Otwock-Świerk, Poland
[3] Faculty of Physics, Astronomy and Applied Computer Science, Jagiellonian University, 30-348 Cracov, Poland
[4] INFN, Laboratori Nazionali di Frascati, 00044 Frascati, Italy
[5] Institute of Physics, Maria Curie-Skłodowska University, 20-031 Lublin, Poland
[6] Faculty of Physics, University of Vienna, 1090 Vienna, Austria
[7] Department of Physics, College of Education for Pure Sciences, University of Mosul, Mosul, Iraq

Science tutor: PhD Krzemień Wojciech
Masełek Rafał: Rafal.Maselek@ncbj.gov.pl

**Keywords: 2+1 photon tomography, J-PET, scandium, positron emission tomography, triple-coincidence**

**Abstract**

    The possibility to separate signals caused by 511 keV photons created in annihilation of electron-positron pairs and the so-called prompt photons from nuclei de-excitation is investigated. It could potentially be used to improve the quality of reconstructed images in the J-PET scanner in 2+1 photon tomography.
    Firstly, a research is conducted for several radioisotopes that decay via β+ decay followed by de-excitation of an excited nucleus. Efficiency, purity and false positive rate are calculated for each isotope as a function of energy deposited threshold, with a hypothesis that signals caused by 511 keV photons deposit smaller values of energy than

the selected threshold, while prompt photons deposit larger energy than the threshold. Analysis of the results accompanied with physical properties of radioisotopes suggests using $^{44}$Sc, which is the most promising candidate for medical applications.

With the use of GATE and J-POS simulation software, in-phantom scattering was introduced and the best energy deposited threshold value was estimated to be approximately 375 keV. It corresponds to almost 100% efficiency for 511 keV signals, 75% purity for 511 keV photons, and approximately 70% efficiency and purity for prompt photons.

## 1. Introduction

### 1.1 Positron Emission Tomography (PET)

Positron Emission Tomography (PET) is a well-established method of medical imaging, used mainly for cancer and many kinds of brain diseases diagnosis. A patient is given a pharmaceutical with selected radioactive isotopes attached. Compounds travel through patient's circulatory system and accumulate in organs, especially in kidneys and bladder. Due to the high metabolism of cancer cells, tracer often accumulates there, providing doctors with a tool for detecting diseases. Radioactive isotopes decay via the β+ process with the emission of positrons — electron's antiparticles. The positron annihilates with an electron from patient's body and their masses are transformed into two back-to-back emitted photons with 511 keV energy each.

Photons leave patient's body and are registered by the PET scanner. Positions and times of registered photons form the input set to reconstruct the map of the tracer activity, which is then used in the diagnosis. Radiologists seek for regions of abnormal emission, which are suspected to contain cancer cells or other malicious processes.

In the first stage of the data reconstruction, the registered signals are filtered according to several criteria to select the events coming from the two 511 keV photons e.g. by selecting pairs of signals within some time period. In the next step, the reconstructed positions of interaction with scintillators are established, and a *line of response (LOR)* is obtained — it is a straight line in space that connects points of interaction. The point of emission is located somewhere on a line of response. The modern PET scanners register not only the position and energy of photons, but also the time of detection in the scintillator. The position of the emission point can be determined more accurately by using the Time-of-Flight method, by measuring the difference in time of registering each of the two signals selected. In principle, if the emission point was located closer to one of the scintillator detectors, then one of the signals will be registered earlier. The closer the source was, the greater is the time interval between registration of two signals.

The registered data set will be contaminated with the events coming from other processes, most notably multiple scattering. There is a high probability that photons created in annihilation process will be scattered inside patient's body before reaching the detector. It may also happen that a photon after scattering in one of scintillator crystals will rescatter in another one. It leads to several misidentification cases:

• At least one of the photons will lose a vast amount of energy, so the algorithm will reject it, as originating from other process than electron-positron pair annihilation.

- At least one of the photons will change its direction, so the line of response will not contain the emission point — in extreme cases the LOR can be located totally outside patient's body.

- One of the photons will not be registered by the detector.

- One of the photons will not be registered by the detector, but another photon (e.g. background originated) will fit into the time window, so an artificial event will be registered.

- One photon will scatter twice in different scintillators, and will be treated as two different signals.

- One of the photons will be correctly registered, however, the second one will scatter twice in scintillator crystals, but only the second scattering will be registered. One of the endpoints of the LOR will be wrong.

Besides of the situations listed above, there many other factors that need to be included when reconstructing an image out of a PET scan. New algorithms are still being developed to decrease the number of fake signals without loosing too much statistics.

**1.2 J-PET experiment**

The J-PET project aims at constructing a novel PET scanner based on plastic scintillators and the dedicated ultra-fast electronics (Moskal et al. 2014; Raczynski et al. 2017; Moskal et al. 2016). Plastic scintillators are much cheaper compared to crystals used in commercial PET scanners. In addition, plastic can be produced in larger blocks, which effectively allows to cover a large part of the patient's body during single measurement. The excellent timing accuracy of about 100 ps (http://koza.if.uj.edu.pl/pet/) permits to use the J-PET scanner also for fundamental studies e.g. discrete symmetries and quantum entanglement (Hiesmayr & Moskal 2017).

In addition to the above mentioned advantages, the J-PET scanner contains many novel improvements when compared to ordinary PET scanners. In typical PET scanners, scintillation crystals are located radially, creating rings that increase the size of the device and allow to measure only small area at once. Plastic scintillation strips of the J-PET scanner are located alongside patient's body. Not only it enables to scan a large body part at once, but also decreases the overall size of the scanner, which opens the possibility to put the J-PET scanner inside other imaging device, such as Magnetic Resonance Imaging scanner or Computer Tomography scanner. Due to the modular construction of the scanner, J-PET can be adjusted to fit into different types of external imaging devices, as well as be adapted for different purposes, for example performing PET scans of large animals. The excellent timing resolution of the J-PET scanner will be further improved by using the silicon multipliers (SIMP).

The current prototype built at the Jagiellonian University laboratory consists of three cylindrical layers of plastic scintillation strips, each with size 7x19x500 mm^3. The inner layer has radius of 425 mm and contains 48 strips, the second one has radius of 467.5 mm and 48 strips, and the outer layer has radius 575 mm and 96 strips. The photo of the prototype is visualized in (Fig.1.). The group at the Jagiellonian University is currently working on constructing a new version of prototype, which would be more compact and have faster electronics.

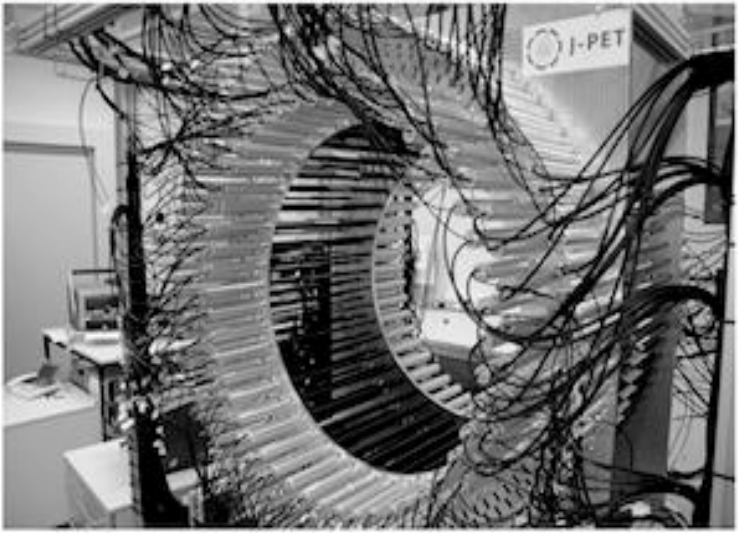

Fig. 1. The first full scale prototype of the J-PET scanner. Platic scintillation strips (black) are visible, as well as photomuliplier tubes (grey) (Niedzwiecki et al. 2017).

**1.3 2+1 photon tomography**

In this study we evaluate the possibility and usefulness of using radioactive isotopes for which positron emission is accompanied by the emission of an additional photon from nucleus de-excitation, the so called *prompt photon*. For such radioisotopes β+ decay can be represented in the following way:

$$^{A}_{Z-1}Y^* \rightarrow ^{A}_{Z-1}Y + \gamma_{prompt},$$

where * superscript denotes excited nucleus. Positron created in a β+ decay annihilates with electron, and two back-to-back photons are emitted. The excited nucleus de-excites:

and a prompt photon is emitted. The energy of the prompt photon is specific for a given isotope. Some radioisotopes can emit prompt photons with one of a few energies, however, in this study only isotopes that emit only one type of prompt photons are taken into account.

In typical PET scanners the events with the registered prompt photons are discarded as unwanted background. In case of the J-PET scanner, the class of three-photon events with the registered prompt photon can be used to improve the image quality. It is based on a critical assumption: the emission of a prompt photons takes place in a short time after the electron-positron annihilation, and nearby. This allows us to treat both processes as a single event consisting of emission of two 511 keV back-to-back photons and a single prompt photon, with the given de-excitation energy, and an isotropic momentum distribution. The registration of three gamma quanta would allow to further reduce the uncertainty of the emission point location, and leads to the drastic improvement in the image accuracy with respect to the two-photon tomography.

## 2. Methods

### 2.1 Simulation framework

All results presented in this study are obtained using a dedicated Monte Carlo simulation framework based on Cern Root 6, called *J-POS (htttps://github.com/JPETTomography/j-pet-ortho-simulations)*. It is optimized for fast simulations of photon production and Compton scattering. The latest version is capable of simulating:

- Single photon emission, with a specific energy, in one of isotropically distributed directions,

- 2-photon emission from positron-electron annihilations,

- 3-photon decays of o-Ps (ortho-positronium) systems,

- 2+1 photon events from beta decays accompanied by nucleus de-excitation. Simulations of radioisotopes, which emit prompt photons with different energies, are possible, however, in this study only those that emit prompt photons with single specific energy are considered.

Smearing of the deposited energy in scintillators, due to finite detectors' efficiency, is implemented using experimentally derived formula:

where E is the energy deposited by incident photon prior to scattering. The energy of the

$$\rho(E) = Gauss(\mu = E,\ \sigma = 0.044\sqrt{E}),$$

scattered photon is randomly generated using the above probability density function. The main source of the background in the PET examination comes from the events in which at least one of the photon is scattered already in the patient's body.

For the purpose of this study, we simulate a simple spherical phantom filled with water. Although, the J-POS simulation framework does not implement full phantom simulation, the change of energy of photons due to initial scattering was considered, assuming that a certain (constant) fraction of all detected photons undergo the scattering in the phantom. The fraction of scattering events is established using GATE simulation software package

### 2.2. Aim of the study

The study is conducted to check how efficiently the prompt de-excitation photon can be distinguished from 511 keV photons based only on the spectra of energy deposited in scintillation strips. Also, the influence of the background coming from the photons scattered in the human body is estimated. Four promising isotopes (Hernandez et al. 2014; Lang et al. 2014) listed in (Tab. 1.) are considered**.**

Tab. 1. Radioisotopes taken into account

| Isotope | $T_{1/2}$ | Energy of prompt photon [keV] | Prompt photon emission probability [%] |
|---|---|---|---|
| $^{14}$O | 71.0 s | 2312 | 99.4 |
| $^{22}$Na | 2.60 y | 1275 | 99.9 |
| $^{44}$Sc | 3.97 h | 1157 | 100 |
| $^{68}$Ga | 67.8 m | 1077 | 1.35 |

If we propose the hypothesis that photons from both processes can be separated by using a single threshold on the deposited energy, then prompt photons are selected by having deposited an energy greater than the threshold, whereas 511 keV lower than the threshold. This suggests using a binary classification for both types of photons.

Tab. 2. Definition of True/False positive/negative terms using a binary classification.

| | | REAL CONDITION | |
|---|---|---|---|
| | | POSITIVE | NEGATIVE |
| PREDICTED CONDITION | POSITIVE | True positive **(TP)** | False positive **(FP)** |
| | NEGATIVE | False negative **(FN)** | True negative **(TN)** |

For this purpose we define the following quantities:

a) True positive rate (TPR/sensitivity/efficiency) — probability of detection:

$$TPR = \frac{TP}{TP + FN}$$

b) Positive predictive value (PPV/purity) — precision rate:

$$PPV = \frac{TP}{TP + FP}$$

c) False positive rate (FPR) — false alarm rate:

$$FPR = \frac{FP}{TN + FP}$$

We show in the following sections how these quantities allow the separation between different processes.

## 3. Results and discussion

### 3.1 Studies of deposited energy spectra of different radioisotopes

The first step of analysis is to compare the deposited energy spectra for selected radioisotopes and to check the possibility to separate signals coming from 511 keV photons from those caused by prompt photons. The process of scattering inside patient's body was not considered. It can be clearly seen from (Fig. 2.) that providing a single threshold on the deposited energy can separate photons in such a way that almost all signals from 511 keV photons would have smaller values than the threshold energy, while the majority of signals from prompt photons would deposit energy greater than the threshold value.

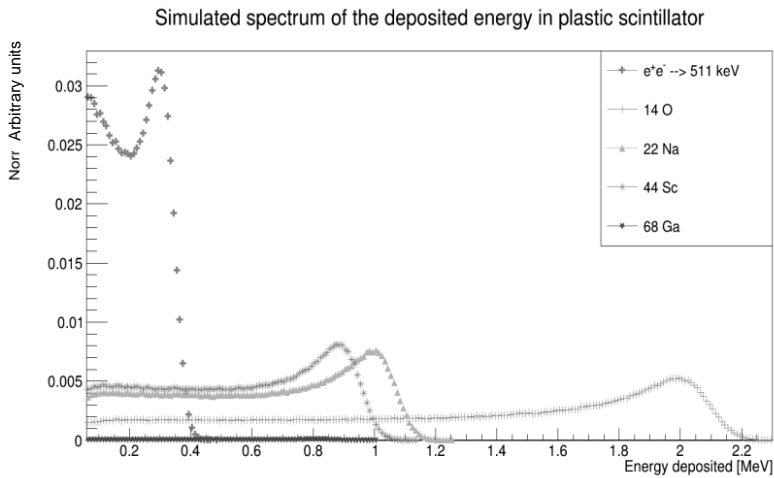

Fig. 2. Spectra of the deposited energy of different radioisotopes. The spectrum of 511 keV is added as a reference. All spectra were smeared according to experimental derived formula and normalized such that the sum of all entries from 511 keV photons add up to one. Spectra for other isotopes were scaled to match the probabilities of prompt photon emission, listed in (Tab. 1.).

To find the optimal value of threshold and to compare different isotopes, two types of plots were plotted in (Fig. 3-6.): purity and efficiency for both types of photons as a function of the energy of threshold, and the so-called ROC (Receiver Operating Characteristics) curve, defined as efficiency vs fall-out value. ROC curves are useful in statistical data analyses. The points that are close to upper left corner of the ROC plot correspond to the best classification (high efficiency and small fall-out).

It can be seen from the ROC plots in (Fig. 3.) that the best separation can be obtained for $^{14}$O radioisotope. For $^{68}$Ga one can obtain, by taking the value of threshold around 425 keV, very good purity for both prompt and 511 keV photons, with almost 100% efficiency for 511 keV photons and an acceptable loss of efficiency for prompts. Results for $^{44}$Sc and $^{22}$Na are similar, but not as good as for $^{14}$O and $^{68}$Ga.

Unfortunately, the probability of prompt photon emission for $^{68}$Ga is very small (c.a. 1%), which makes it non-applicable for 2+1 photon tomography. Similarly, short half-life time for $^{14}$O (71s) discards it. Out of the two isotopes left, $^{44}$Sc is more suitable for medical use due to half-life time of approximately 4h, compared to over 2 years for $^{22}$Na.

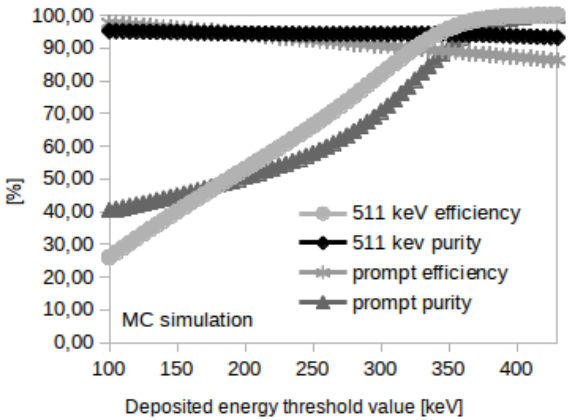 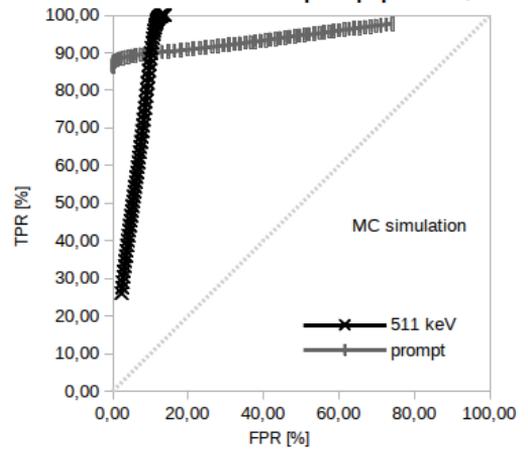

Fig. 3. Efficiency and purity as a function of deposited energy threshold value (left) and ROC curves (right) for $^{14}$O.

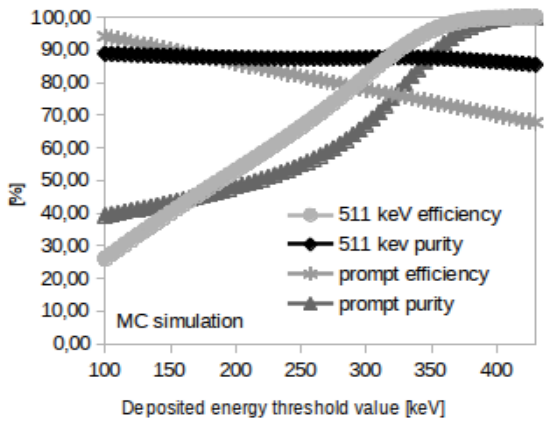 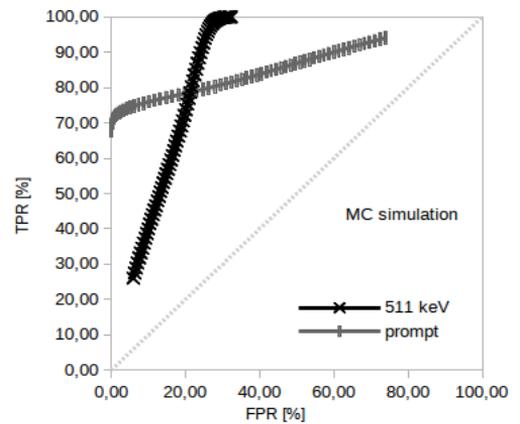

Fig. 4. Efficiency and purity as a function of deposited energy threshold value (left) and ROC curves (right) for $^{22}$Na.

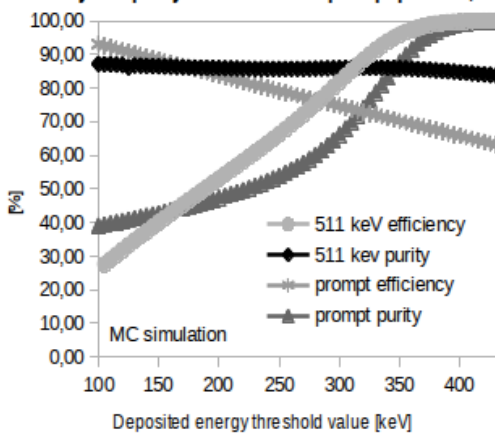 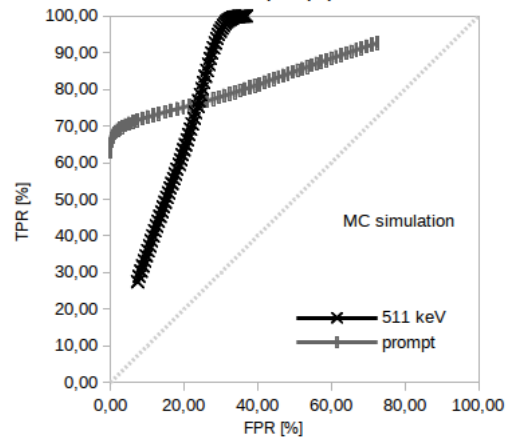

Fig. 5. Efficiency and purity as a function of deposited energy threshold value (left) and ROC curves (right) for $^{44}$Sc.

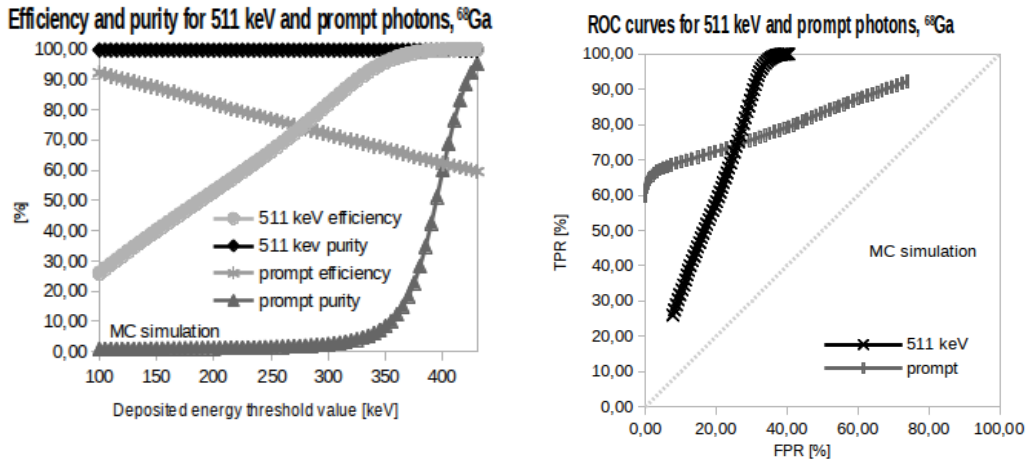

Fig. 6. Efficiency and purity as a function of deposited energy threshold value (left) and ROC curves (right) for $^{68}$Ga. The efficiency for 511 keV signals is almost 100%, because of the very low emission probability of prompt photon for $^{68}$Ga, resulting in low number of prompt photons compared to the number of photons created in annihilation processes.

Among the four radioisotopes, the performed studies show that $^{44}$Sc is the most promising candidate for the medical 2+1 photon tomography. The research presented in the following sections of this article is conducted only for $^{44}$Sc radioisotope.

### 3.2 Determination of the fraction of scattered events in the body phantom for $^{44}$Sc

In a realistic medical examination some fraction of photons, produced in electron-positron annihilation and nucleus de-excitation processes, is scattered inside patient's body. These photons cannot be used to reconstruct the location of the emission point, and create an unwanted background. In this study a rough approximation of such background is made, by adding to signal spectra of 511 keV and prompt photons after a single Compton scattering process. These spectra are weighted using ratios of numbers of in-phantom scattered photons to all registered. Ratios were obtained using Gate simulation software (http://www.opengatecollaboration.org) for ball-shaped water-based phantom, for four different radii. The results are presented in (Fig. 7.) together with theoretical formula for linear absorption, given by:

$$ratio = 1 - e^{-\mu z}$$

One can see from (Fig. 7. left part) that ratios obtained using GATE simulations have greater values than theoretical predictions, however, the general characteristic of both distributions seems similar. Indeed, one can easily find, that results of the simulation follow the same formula, but with absorption factors scaled according to:

$$\mu_{sim} = 1.21 \cdot \mu_{theory}$$

One can see in (Fig. 7. Right part) that after scaling the theoretical absorption coefficient, simulated data match theoretical prediction perfectly.

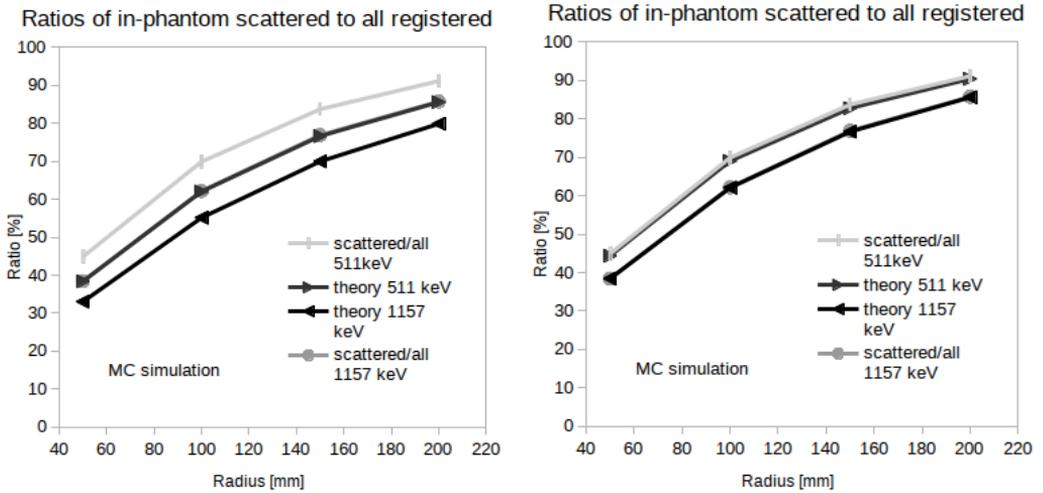

Fig. 7. Ratios of numbers of in-phantom scattered photons to all registered for 511 keV and prompt photons for $^{44}$Sc. Theoretical estimations for linear absorption process are also included. The right image presents curves for which absorption coefficients are scaled according to formula $\mu_{sim} = 1.21 \nu_{theory}$.

### 3.3 Deposited energy spectrum for $^{44}$Sc with in-phantom scattering

The spectrum of energy deposited for $^{44}$Sc is presented in (Fig.8.)
A background resulting from 511 keV is included, in addition to prompt photons that were initially scattered inside phantom with radius equal to 100 mm.

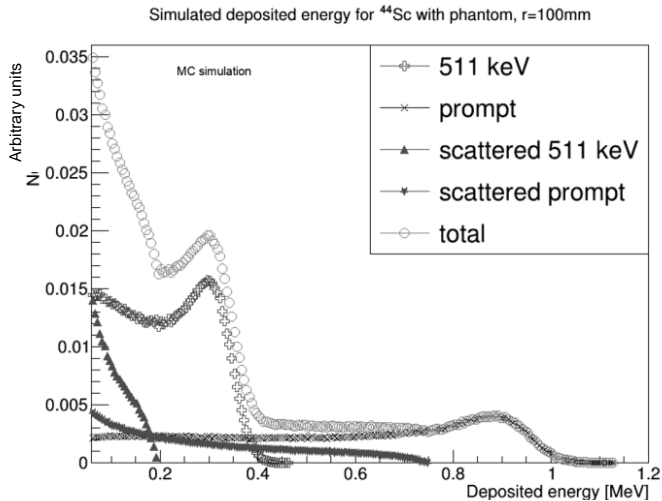

Fig. 8. Energy deposited spectrum for 44Sc with phantom, with radius 100 mm. Constrictions from 511 keV and 1157 keV photons are presented, both in-phantom scattered and unscattered signals. The total (summed) spectrum is also provided.

Simulated deposited energy spectra for phantoms with radius of 50 mm, 150 mm and 200 mm look very similar. One can see from (Fig. 8.) that the background from scattered 511 keV can be removed by applying a single selection and disregarding signals with deposited energy lower than 200 keV. In contrary, the contribution from in-phantom scattered prompt photons cannot be removed in such way. However, one can safely assume that above approximately 750 keV all signals come from non-scattered prompt photons. Therefore, threshold should be located somewhere between 200 keV and 750 keV as it is presented in (Fig. 9.).

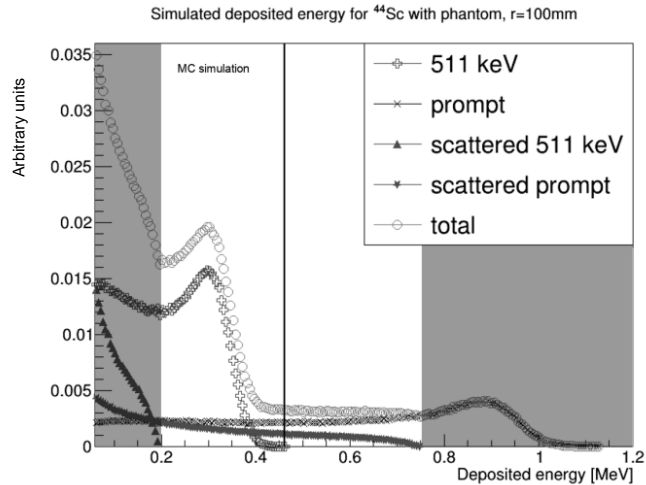

Fig. 9. Energy deposited spectrum for 44Sc with phantom, with radius 100 mm. Example value of threshold energy is denoted by the vertical line. The interesting region for threshold optimization is 200-750 keV.

### 3.4 Optimal threshold value

To estimate the optimal deposited energy value for threshold, plots of efficiency, purity and ROC curves are plotted, similarly to section 3.1. They are presented in (Fig. 10.) and (Fig. 11.). The plots presented in (Fig. 10.) and (Fig. 11.) differ from plots presented in section 3.1 for $^{44}$Sc, because they include background from prompt photons scattered in the phantom (background from in-phantom scattered 511 keV photons is removed by selecting only signals with deposited energy value greater than 200 keV). One of the possible criteria for electing the best classification is to choose the threshold value corresponding to the cross-section of ROC curves for prompt and 511 keV photons. The value of deposited energy threshold for that point is about 325 keV and corresponds to the intersection of efficiency curves in (Fig. 10.).

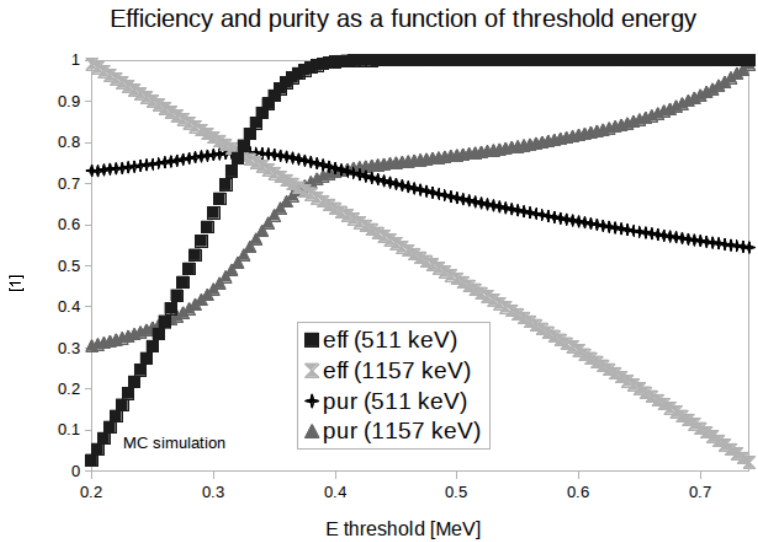

Fig. 10. Efficiency and purity for [44]Sc with a water phantom with radius of 100 mm.

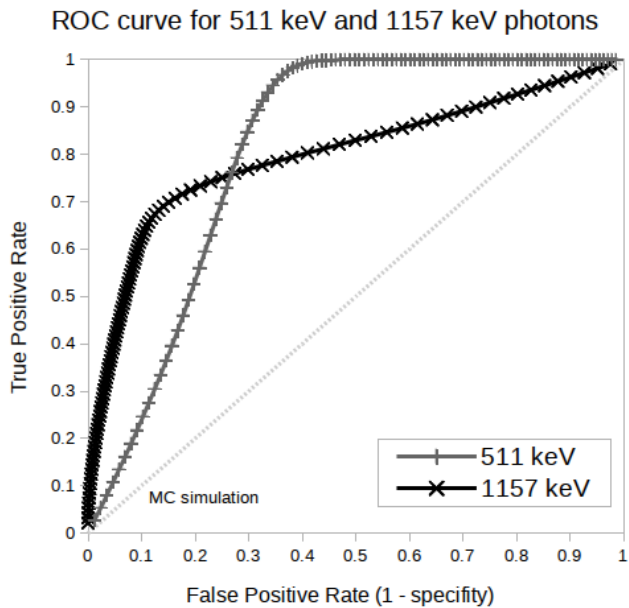

Fig. 11. ROC curve plot for 44Sc with a water phantom with radius of 100 mm used.

Other criteria, is to treat signals from 511 keV photons as the main source of the information, while signals caused by prompt photons as an additional one. It corresponds to maximizing efficiency for 511 keV photons, and thus choosing for instance the threshold value around 375 keV, for which purity and efficiency curves of 1157 keV photons intersect.

### 4. Conclusions

Results of the study confirm that [44]Sc isotope is a good candidate for the 2+1 photon tomography. It is possible to efficiently distinguish between 511 keV and high-energetic 1157 keV photons based only on the spectrum of deposited energy in plastic scintillation strips, with deposited energy threshold set to ca. 375 keV. It enables to achieve almost 100% efficiency and 75% purity for 511 keV photons, while obtaining approximately 70% efficiency and purity for prompt photons. The half-life time for $^{44}$Sc is approximately 4 hours, which is enough to transport radioisotopic materials to hospitals and health centers and to conduct medical examination. The contribution to energy deposited spectra from phantom-scattered 511 keV photons can be removed by simply selecting only signals with deposited energy larger than 0.2 MeV. The contribution from in-phantom scattered prompt photons cannot be removed in a similar manner. They form a linear background contributing by a factor of 13% to the overall statistics.

Due to the excellent J-PET angular resolution (~ 1 degree), the reconstructed information about the photon detection position can be used to construct statistically independent method based on the angular dependence of detected photons, that can be used to further reduce the background coming from the scattering prompt photons. The